\documentstyle[12pt]{article}
\newcommand{\h}[1]{\mathop{\lambda}\limits_{#1}\ \!\!\!}

\newcommand{\ps}{\hspace{.3cm}}
\newcommand{\p}{\hspace{.5cm}}
\newcommand{\pp}{\hspace{1cm}}
\newcommand{\pppp}{\hspace{2cm}}
\newcommand{\ppppp}{\hspace{3cm}}

\newcommand{\ppp}{\hspace{1.5cm}}

\begin{document}

\title
{ Effect of Spin-Torsion Interaction on Raychaudhuri Equation }
\author{ M.I.Wanas \thanks{Astronomy Department,Faculty of Science,Cairo University,
Giza,Egypt
e.mail:wanas @ frcu.eun.eg}
\and
M.A.Bakry
\thanks{Mathematics Department,Faculty of Education,Ain Shams University,Cairo,Egypt.}}
\date{}
%\begin{document}
\newpage
\maketitle
\section*{Abstract}
Raychaudhuri equation is generalized in the parameterized absolute
parallelism geometry. This version of absolute parallelism  is more
general than the conventional one. The generalization takes into
account the suggested interaction between the quantum spin of the
moving elementary particle and the torsion of the background
gravitational field. The generalized Raychaudhuri equation obtained
contains some extra terms, depending on the torsion of space-time,
that would have some effects on the singularity theorems of general
relativity. Under a certain condition, this equation could be
reduced to the original
Raychaudhuri equation without any need for a vanishing torsion.\\
\begin{center}
{\bf KEY WORDS: Singularity - Absolute Parallelism Geometry- Path
Equations, Anti-gravity, Torsion}
\end{center}
\newpage
\section{Introduction}
It is well known that the Raychaudhuri equation plays an essential
role in the study of space-time singularities [1]. Singularity
theorems, established using this equation, show that the existence
of singularities, in the solutions of general relativity (GR), is
inevitable. Several attempts have been done to generalize or modify
the Raychaudhuri equation in the hope that GR or other geometric
field theories will be free from such singularities (cf.[2]).
However, Senovilla [3]
 obtained some solutions of the field equations of GR which are
not singular, by relaxing some of the assumptions of the singularity
theorems of GR. Thus it is necessary to examine the roots of the
singularity theorems, i.e. the Rayshaudhuri equation. This may give
rise to other factors that affect the existence of singularities
without any need to relax the assumptions mentioned above.

It is widely accepted that a collapsed object, before approaching a
singular state, passes through a state in which matter degenerates
into its elementary constituents. GR assumes that motion of
elementary particles in a gravitational field is along geodesics of
the metric, regardless of the spin of these particles. On the other
hand, Raychaudhuri equation depends on the validity of the geodesic
motion of these particles. If there is an interaction between the
spin of the moving elementary particle and the background
gravitational field, then there will be deviation from the geodesic
motion and consequently the Raychaudhuri equation is to be modified
to take this interaction into account.

One of the authors [4] suggested a version of {\it "Absolute
Parallelism"} (AP)-Spaces in which curvature and torsion are
simultaneously non-vanishing objects. This type of structure is
known as the {\it "Parameterized Absolute Parallelism"} (PAP) space.
Among other things, the non-symmetric connection of the AP-geometry
is generalized, and consequently a new path equation is derived.
This equation is suggested to represent trajectories of spinning
particle in gravitational fields. The equation contains a term, that
is suggested to represent a type of interaction between the quantum
spin of the moving  particle and the background gravitational field.
This interaction has been called {\it "spin-gravity interaction" or
"spin-torsion interaction"}. The equation, in its linearized form,
has been used to interpret the discrepancy appeared in the results
of the COW-experiment [5], and to discuss the time delay of massless
particles coming from SN1987A [6].

The aim of the present work is to generalize the Raychaudhuri
equation using the new general affine connection and the new path
equation, in place of the geodesic of the metric. In section 2 a
brief account on the recently established structure of the AP-space
is given. Generalization of the Raychaudhuri equation with necessary
definitions are given in section 3. Discussion of the results
obtained and some remarks are given in section 4.
\section{The Parameterized AP-geometry}
The structure of the conventional AP-space is defined completely, in
4-dimensions,  by a tetrad vector $\h{i}^\mu$($i(=1, 2, 3, 4)$
stands for the vector number and $\mu(=1, 2, 3, 4)$ stands for the
coordinate component) which is subject to the AP-condition,
\begin{eqnarray}
\h{i}^{_{\stackrel{\bf{\alpha}}{+}}}_{~|{~\beta}}  =   0~~~.
\end{eqnarray}
The covariant components of $\h{i}^\alpha$ is defined such that:
\begin{eqnarray}
\h{i}^\alpha  \h{i}_{\mu} = \delta^{\alpha}_{\mu}.
\end{eqnarray}
Summation convention is carried out over repeated indices whatever
their position. The condition (1) defines a non-symmetric connection
$\Gamma^\alpha_{. \mu \nu} ( \stackrel{def.}{=} \h{i}^\alpha
\h{i}_{\mu ,\nu})$ . A second order tensor, which can play the role
of the metric tensor,is defined by,
\begin{eqnarray}
g_{\alpha \beta}  \stackrel{def.}{=}  \h{i}_{\alpha}  \h{i}_{\beta}~~~.
\end{eqnarray}
The torsion of this space is defined by the tensor,
\begin{eqnarray}
\Lambda^\alpha_{. \mu \nu} \ps  \stackrel{def.}{=} \ps\Gamma^\alpha_{. \mu \nu}
- \Gamma^\alpha_{. \nu \mu } \p .
\end{eqnarray}

Certain considerations [4] lead to the following general linear
connection,
\begin{eqnarray}
\nabla^\mu_{. \alpha \beta} \ps \stackrel{def.}{=} \ps a\left\{\begin{array}{c}
\mu\\ \alpha\beta\end{array}\right\}+b \Gamma^{\mu}_{ \alpha\beta}
\end{eqnarray}
where $a,b$  are parameters and $\left\{\begin{array}{c} \mu\\
\alpha\beta\end{array}\right\}$ is the Christoffel symbol defined
using (3).
 The tensor derivative using (5) is defined by,
\begin{eqnarray}
A_{\mu || \nu} \ps \stackrel{def.}{=}\ps A_{. \mu,\nu} - A_\alpha \nabla^\alpha_{. \mu \nu} \p ,
\end{eqnarray}
\[ A^\mu _{~|| \nu} \ps \stackrel{def.}{=}\ps A^\mu_{ ,\nu} + A^\alpha \nabla^\mu_{. \alpha \nu}\p , \]
where $A_\mu$ is an arbitrary vector. Metricity is achieved upon taking the condition,
\begin{eqnarray}
a + b \ps = \ps 1,
\end{eqnarray}
\begin{eqnarray}
\mbox{i.e.} \pp  g_{\mu \nu|| \sigma} \ps = \ps 0.
\end{eqnarray}
The non-commutation of the general tensor derivatives is expressed
by [4]
\begin{eqnarray}
A_{\mu || \nu \sigma} - A_{\mu || \sigma \nu} \ps
= \ps A_\alpha B^{\alpha}_{. \mu \nu \sigma} - b A_{\mu || \varepsilon} \Lambda^\varepsilon_{. \nu \sigma}
\end{eqnarray}
where,
\begin{eqnarray}
B^{\alpha}_{. \mu \nu \sigma} \ps \stackrel{def.}{=} \ps \nabla^\alpha_{. \mu \sigma,\nu}
- \nabla^\alpha_{. \mu \nu,\sigma} + \nabla^\alpha_{. \varepsilon \nu} \nabla^\varepsilon_{. \mu \sigma}
- \nabla^\alpha_{. \varepsilon \sigma} \nabla^\varepsilon_{. \mu \nu}\ps .
\end{eqnarray}
Using the general affine connection (5) and the metricity condition
(8),the following path equation is obtained [4],
\begin{eqnarray}
\frac{dZ^\mu}{d\tau} + \left\{ \begin{array}{c} \mu \\ \alpha \beta
\end{array} \right\} Z^\alpha Z^\beta \ps = \ps - b \Lambda^{ \hspace{.4cm} \mu}_{( \alpha \beta)} Z^\alpha Z^\beta \p ,
\end{eqnarray}
where parantheses are used for symmetrization, $Z^\mu
(\stackrel{def.}{=} \frac{dx^\mu}{d\tau})$ is the tangent to the
path and $\tau$ is the parameter varying along the path. The
parameter $(b)$ is suggested to take the form [4],
\begin{eqnarray}
b \ps = \ps \frac{n}{2} \alpha \gamma \p ,
\end{eqnarray}
where $n$ is a natural number, $\alpha$ is the fine structure
constant and $\gamma$ is a dimensionless parameter to be fixed by
experiment or observation. The term on the R.H.S. of (11) is
suggested [4] to represent a type of interaction between the
intrinsic quantum spin, of the moving  particle and the torsion of
the background gravitational field. The natural number $n$ of (12)
takes the values $0,1,2,\ldots$ for particles with spin
$0,\frac{1}{2},1,\ldots$ respectively.\\ Equation (11) is suggested
to represent the motion of spinning elementary particle in a
gravitational field. It is worth of mention that for spinless
particles and slowly rotating macroscopic bodies $n = 0$, and thus
equation (11) will reduce to the ordinary geodesic equation.\\ An
important note is that in the case $a = 1 , b = 0 ,$ the affine
connexion (5) reduces to Christoffel symbol and the geometry becomes
Riemannian, while in the case $a = 0 , b = 1,$ the connexion (5)
reduces to $\Gamma^\alpha_{. \mu  \nu}$ and the geometry becomes the
conventional AP-geometry. The geometry, depending on the connection
(5) and the condition (7), given briefly in this section is called
the {\it "Parameterized Absolute Parallelism"} (PAP) geometry. For
more details, the reader is referred to reference [7].
%\newpage
\section{Generalization of Raychaudhuri scheme}
In this section, we are going to generalize the original
Raychaudhuri scheme [8], in order to study the effect of {\it
spin-torsion interaction} on the structure of this equation.
According to the PAP-geometry, reviewed briefly in section 2, the
velocity components of a spinning particle, assumed to move along
(11), in a gravitational field is given by,
\begin{eqnarray}
Z^\mu  \ps \stackrel{def.}{=}\ps \frac{dx^\mu}{d\tau} \p .
\end{eqnarray}
Using this vector,we can define the following tensors:
\begin{eqnarray}
Z_{\mu \nu} \ps \stackrel{def.}{=}\ps Z_{\mu || \nu} \p ,
\end{eqnarray}
\begin{eqnarray}
\Theta \ps \stackrel{def.}{=}\ps Z^\mu _{~~|| \mu} \p ,
\end{eqnarray}
\begin{eqnarray}
\Omega_{\mu \nu} \ps \stackrel{def.}{=}\ps Z_{[\mu || \nu]} \p ,
\end{eqnarray}
\begin{eqnarray}
\Sigma_{\mu \nu} \ps \stackrel{def.}{=}\ps Z_{(\mu || \nu)} - \frac{1}{3} \frac{\Theta}{Z^2} P_{\mu \nu}\p ,
\end{eqnarray}
where brackets are used for anti-symmetrization and,
\begin{eqnarray}
P_{\mu \nu} \ps \stackrel{def.}{=}\ps Z^2 g_{\mu \nu} -Z_\mu Z_\nu \p ,
\end{eqnarray}
and
\begin{eqnarray}
Z^2 \ps \stackrel{def.}{=}\ps Z^\mu Z_\mu \p .
\end{eqnarray}

The tensors $\Omega_{\mu \nu} , \Sigma_{\mu \nu}$ are generalization
of the vorticity and shear tensors, in Raychaudhuri treatment
respectively, while the scalar $\Theta$ is a generalization of the
expansion.The tensor $P_{\mu \nu} $ can be used as a generalization
of the projection tensor, as it projects any tensor on the
hypersurface perpendicular to the vector $Z^\mu$ as will appear in
due course.

 It can be easily shown
that,
\begin{eqnarray}
P^{. \nu}_\mu P^\mu_{. \nu} \ps = \ps 3 Z^4\p ,
\end{eqnarray}
\begin{eqnarray}
P^\mu_{. \mu} \ps = \ps 3 Z^2 \p .
\end{eqnarray}
The non-symmetric tensor $Z_{\mu \nu} $ can now be written in the form:
\begin{eqnarray}
Z_{\mu \nu} \ps = \ps  \Omega_{\mu \nu} + \Sigma_{\mu \nu} + \frac{1}{3} \frac{\Theta}{Z^2} P_{\mu \nu}\p .
\end{eqnarray}
It has been shown [5], using the 1st integral of the path equation
(11), that $Z$ is constant, thus
\[ \left( Z^\beta Z_\beta \right)_{|| \alpha} \ps = \ps 0 \p .\]
As a consequence of (8),the last expression can be written as,
\[ Z^\beta Z_{\beta || \alpha} \ps = \ps 0 \ps ,\ppppp (i)\]
while the new path equation (11) can be written in the equivalent
form,
\[ Z^\beta Z^{\alpha}_{. || \beta} \ps = \ps 0 \ps .\ppppp (ii)\]
Now using {\it{(i)}} , {\it{(ii)}}, the following results can be
easily obtained:
\begin{eqnarray}
\Sigma_{\alpha \beta} Z^\beta \ps = \ps 0 \p ,
\end{eqnarray}
\begin{eqnarray}
\Omega_{\alpha \beta} Z^\beta \ps = \ps 0 \p ,
\end{eqnarray}
from which it is clear that $\Sigma_{\alpha \beta} , \Omega_{\alpha
\beta}$ are defined only on the hypersurface perpendicular to the
$Z^\mu$ vector, i.e. they are purely spatial tensors. Hence we can
verify the following results:
\begin{eqnarray}
P^{\alpha \beta} \Sigma_{\alpha \beta}  \ps = \ps 0 \p ,
\end{eqnarray}
\begin{eqnarray}
P^{\alpha \beta}\Omega_{\alpha \beta}  \ps = \ps 0 \p .
\end{eqnarray}
Let us now define the following quantities,
\begin{eqnarray}
\Omega^2 \ps \stackrel{def.}{=}\ps  \frac{1}{2} \Omega_{\alpha \beta}  \Omega^{\alpha \beta} \p ,
\end{eqnarray}
\begin{eqnarray}
\Sigma^2 \ps \stackrel{def.}{=}\ps  \frac{1}{2} \Sigma_{\alpha \beta}  \Sigma^{\alpha \beta} \p .
\end{eqnarray}
Using (19), (24), (25), (26) and (27), and noting that
$\Omega_{\alpha \beta}$ is antisymmetric, we can write
\begin{eqnarray}
Z^{\alpha}_{.\beta}Z^{\beta}_{.\alpha}=-2\Omega^{2}+2\Sigma^{2}+\frac{1}{3}\Theta^{2}
\end{eqnarray}
From the new path equation {\it{(ii)}} we can write

\[ \left(Z^\alpha Z^\beta_{.~|| \alpha} \right)_{|| \gamma} \ps = \ps 0 \p ,\]
i.e.  \[ Z^\alpha_{.~|| \gamma}Z^\beta_{.~|| \alpha} + Z^\alpha
Z^\beta_{.~|| \alpha \gamma} \ps = \ps 0 \p . \pppp (iii)\] Using
(9), we can write,
\[Z^\beta_{.~|| \alpha \gamma} \ps  = \ps Z^\sigma B^{.\beta}_{\sigma . \alpha \gamma} +
Z^\beta_{.~|| \gamma \alpha} - b Z^\beta_{.~|| \sigma} \Lambda^\sigma_{. \alpha \gamma}  \p . \ppp (iv)\]
Substituting from {\it{(iv)}} into {\it{(iii)}}  we get,
\[ Z^\alpha_{.~|| \gamma} Z^\beta_{.~|| \alpha} + Z^\alpha (Z^\sigma B^{.\beta}_{\sigma . \alpha \gamma} +
Z^\beta_{.~|| \gamma \alpha} - b Z^\beta_{.~|| \sigma}
\Lambda^\sigma_{. \alpha \gamma})  \ps = \ps 0  \ps .\] Contracting
the last equation by setting $\beta= \gamma$, we get after using
(15), (27) and (28),
\begin{eqnarray}
\frac{d \Theta}{d \tau} \ps = \ps 2 \Omega^2 - 2 \Sigma^2 -\frac{1}{3} \Theta^2
- Z^\alpha Z^\sigma B_{\sigma \alpha} + b Z^\alpha Z^\beta_{.~|| \sigma} \Lambda^\sigma_{. \alpha \beta} \p .
\end{eqnarray}
where,
\begin{eqnarray}
B_{\sigma \alpha} \ps \stackrel{def.}{=}\ps B^{.\beta}_{\sigma . \alpha \beta} \p .
\end{eqnarray}
Substituting from (16) and (17) into (30),we get
\begin{eqnarray}
\frac{d \Theta}{d \tau} \ps = \ps 2 \Omega^2 - 2 \Sigma^2
-\frac{1}{3} \Theta^2 - Z^\alpha Z^\sigma B_{\sigma \alpha} + b
Z^\alpha (\Omega^\beta_.\sigma + \Sigma^\beta_.\sigma)
\Lambda^\sigma_{. \alpha \beta} + b \frac{\Theta}{3} Z^\alpha
C_\alpha \p .
\end{eqnarray}
\[ \mbox{where} \pp C_\alpha \ps = \ps \Lambda^\beta_{. \alpha \beta}\pp \mbox{and} \pp Z_\sigma Z^\alpha Z^\beta
\Lambda^\sigma_{. \alpha \beta} \ps = \ps 0 \ps . \ppp\] Equation
(32) is a generalized form of Raychaudhuri equation in which the
suggested interaction, between the quantum spin of the moving
elementary particle and the torsion of the background gravitational
field, is taken into account.
\section{Discussion}
In the PAP-geometry, (having a general non-symmetric connection, and
simultaneously non-vanishing torsion and curvature), Raychaudhuri
equation is generalized. The generalization takes into account the
suggested interaction between the quantum spin of the moving
particle and the background space-time torsion produced by a
gravitational field.

If we compare the generalized equation (32) with the original
Raychaudhuri equation, developed in Riemannian geometry [8], we find
some extra terms. Some of those terms appeared as a consequence of
the generalized definitions (13)-(19), i.e. due to the use of the
general affine connection (5) with the condition (7). These terms
are hidden in all the terms of (32) except the last two terms in
which $(b)$ appears explicitly. In general all these terms depend on
the new parameter $b$. The vanishing of this parameter will reduce
(32) to the original Raychaudhuri equation. The vanishing of $b$
indicates that the interaction between quantum spin and torsion is
switched off. In other words, the vanishing of the parameter $b$
indicates that the moving particle is spinless.

Equation (32) can be compared with a corresponding one established
in spaces with torsion [2]. For example, in the Riemann-Cartan space
in which Einstein-Cartan theory is constructed, the torsion is
coupled with the intrinsic spin density of the material particles,
which vanishes in the absence of spin [9]. In this case we encounter
a problem, that is, such theories are not viable in conventional
AP-spaces [10], since the vanishing of the torsion will reduce the
space-time to a flat one. In the present treatment, torsion is not
connected to the spin density, but to the gravitational field. For
gravity theories constructed in the AP-space (cf.[11]), (even in the
case of GR written in this space) the space time torsion has
non-vanishing components whether or not the spin is present.

The present treatment overcomes the above mentioned problem. The
R.H.S of equation (11) represents, as stated before, a type of
interaction between the quantum spin and the torsion of the
background gravitational field. There is no need for the torsion to
vanish in order to switch off this interaction. The interaction will
vanish automatically for spinless particles ($n = 0$), in virtue of
(12). More investigation is needed to explore the effect of the
extra term on the singularity theorems.

Finally, we would like to point out that, the interaction of matter
with space-time curvature gives rise to an {\bf attractive} force
(gravity), while its  interaction with space-time torsion gives rise
to a {\bf repulsive} one (anti-gravity)[4]. This behavior has been
used to interpret the accelerating expansion of the Universe [12].
\subsubsection*{Acknowledgments}
The authors would like to thank late Dr. M.Melek, and Dr. M.E.Kahil
for many discussions.
\subsubsection*{References}
\begin{itemize}
%\begin{enumerate}
\item [1] Hawking,S.W. and Ellis,G.R.F. (1973) { \it Large scale structure of space-time},
      ${\  }$ Cambridge University Press.
\item [2] Tafel,J.(1973) Physics Letters  {\bf {45A}}, 341.\\
      Raychaudhuri, A.K.(1979) {\it Theoretical cosmology}, Springer Verlag.\\
      Garcia de Andrade,L.C.(1990) Int. J.  Theoret. Phys.
      ${\ }$  {\bf{29}}, 997.
\item [3] Senovilla, Jos{\'e} M.N. (1990) Phys. Rev. Let., {\bf{ 64}}, 2219.\\
      Chinea, F.J., Fernandez-Jambrina, L. and Senovilla, Jos{\'e}
      M.N. (1992) Phys. Rev. D, {\bf 45}, 481.\\
      Senovilla, Jos{\'e} M.N. (1996) Phys. Rev. D, {\bf {53}}, 1799.
\item [4] Wanas, M.I.(1998) Astrophys. Space Sci.,{\bf
{258}}, 237. gr-qc/9904019.\\ Wanas, M.I. (2000) Turk.J.Phys. {\bf
24}, 473. gr-qc/0010099.
\item [5] Wanas,M.I., Melek,M. and Kahil,M.E. (2000) Gravitation and Cosmology {\bf 6}, 319. gr-qc/9812085.
\item [6] Wanas, M.I., Melek, M. and Kahil, M.E. (2002) Proceedings of MG IX Vol.B,
1100. gr-qc/0306086.
\item [7] Wanas, M.I. (2002) Proceedings of MG IX Part B, 1303.\\
          Wanas, M.I. (2001) Proceedings of the 11th conference on
          {\it Finsler, Lagrange and Hamilton
          Geometries}.gr-qc/0209050.

\item [8] Raychaudhuri,A.K. (1955) Phys. Rev. {\bf {98}}, 1123.
\item [9] Raychaudhuri,A.K. (1975) Phys. Rev. D,{\bf {12}}, 952.
\item [10] Wanas,M.I. and Melek,M. (1995) Astrophys.  Space Sci., {\bf {228}}, 277.
\item [11] Mikhail,M.I. and Wanas,M.I. (1977) Proc. Roy. Soc. London, {\bf A 356}, 471. \\
      Moller,C.(1978) Mat.-Fys. Skr.  Dans. Vid. Selsk., {\bf {39}}, 1.\\
      Wanas,M.I.(1981) Nuovo Cimento, {\bf {66}}, 145.
\item [12]  Wanas,M.I. (2007) arXiv: 0704.3760. \\
      Wanas,M.I. (2007) Int. J. Mod. Phys. A, {\bf {22}} (31),
      5709;arXiv:08o2.4104
%\end{enumerate}
\end{itemize}

\end{document}